\documentclass[journal]{IEEEtran}
\usepackage{xcolor,soul,framed} 
\colorlet{shadecolor}{yellow}
\usepackage[pdftex]{graphicx}
\usepackage{caption}
\usepackage{subcaption}
\graphicspath{{../pdf/}{../jpeg/}}
\DeclareGraphicsExtensions{.pdf,.jpeg,.png}
\usepackage[cmex10]{amsmath}
\usepackage{amssymb}
\usepackage{amsthm}
\usepackage{amsmath}
\usepackage{array}
\usepackage{mdwmath}
\usepackage{mdwtab}
\usepackage{eqparbox}
\usepackage{url}
\usepackage{tabularx}
\usepackage{booktabs}
\begin{document}
\bstctlcite{IEEEexample:BSTcontrol}
    \title{Ground-to-UAV Integrated Network: Low Latency Communication over Interference Channel}
  \author{Sudhanshu~Arya,~\IEEEmembership{Member,~IEEE}
      and~Ying~Wang\,~\IEEEmembership{Member,~IEEE}

  \thanks{(Corresponding author: Ying Wang.)}
  \thanks{Sudhanshu Arya and Ying Wang are with the School of Systems and Enterprises, Stevens Institute of Technology, Hoboken, USA (e-mail: sarya@stevens.edu; ywang6@stevens.edu).}
  }


\maketitle

\begin{abstract}
We present a novel and first-of-its-kind information-theoretic framework for the key design consideration and implementation of a ground-to-UAV (G2U) communication network with an aim to minimize end-to-end transmission delay in the presence of interference. The proposed framework is useful as it describes the minimum transmission latency for an uplink ground-to-UAV communication must satisfy while achieving a given level of reliability. To characterize the transmission delay, we utilize Fano’s inequality and derive the tight upper bound for the capacity for the G2U uplink channel in the presence of interference, noise, and potential jamming. Subsequently, given the reliability constraint, the error exponent is obtained for the given channel. In addition, as a function of the location information of the UAV, a tight lower bound on the transmit power is obtained subject to the reliability constraint and the maximum delay threshold. Furthermore, a relay UAV in the dual-hop relay mode, with amplify-and-forward (AF) protocol, is considered, for which we jointly obtain the optimal positions of the relay and the receiver UAVs in the presence of interference, and is compared with the point-to-point G2U link which ignores the relay. Interestingly, in our study, we find that, for both the point-to-point and relayed links, increasing the transmit power may not always be an optimal solution for delay minimization problems. In particular, we show that increasing the power gives a negligible gain in terms of delay minimization, though may greatly enhance the outage performance. Moreover, we prove that there exists an optimal height that minimizes the end-to-end transmission delay in the presence of interference. The proposed framework can be used in practice by a network controller as a system parameters selection criteria, where among a set of parameters, the parameters leading to the lowest transmission latency can be incorporated into the transmission. The based analysis further set the baseline assessment when applying Command and Control (C2) standards to mission-critical G2U and UAV-to-UAV (U2U) services. 
\end{abstract}


\IEEEpeerreviewmaketitle

\section{Introduction}
\subsection{Background and Motivation}
\IEEEPARstart{5G}{} and beyond wireless systems are expected to support an ultra-reliable and low-latency communication link (URLLC) to enable uninterrupted and ubiquitous connectivity to mission-critical services where robust information exchange is important \cite{ma2019high}. Indeed, the third generation partnership project (3GPP) aims to cover three generic connectivity technologies for 5G and beyond systems: URLLC, enhanced mobile broadband (eMBB), and massive machine-type communication (mMTC) \cite{popovski2019wireless}. eMBB is designed for applications that have high requirements for the data rate, such as high-resolution video streaming whereas mMTC focuses on applications that support the massive number of machine-type devices with ultra-low power consumption. In contrast, URLLC is designed for delay-sensitive services and applications. For example, the services including intelligent transportation systems and tactile internet have reliability requirements of $(1 - 10^{-3})  \sim  (1 - 10^{-9})$ at latency between $1$ ms to $100$ ms \cite{shirvanimoghaddam2018short}. However, due to the dynamic and fast-fading channel, shadowing, and path loss over wireless links, it becomes challenging to meet the quality-of-service (QoS) requirement for URLLC systems.

To meet the challenging demands of 5G wireless networks, 3GPP has recommended integrating unmanned aerial vehicles (UAVs) into wireless cellular networks \cite{muruganathan2021overview}. With a high probability of establishing a LOS communication link, UAVs are more likely to provide better link quality over short-distance. Moreover, UAVs are generally cost-effective and offer more flexibility for on-demand communication systems in low-altitude environments. As a result, UAVs have received significant research attention in wireless communications \cite{zeng2018cellular,lin2018sky}.

A system model for UAV relay-assisted IoT networks was presented to primarily explored the impact of requested timeout constraints for uplink and downlink transmissions \cite{tran2021uav}. A problem to maximize the total number of served IoT devices was formulated. The authors jointly optimized the transmission power, trajectory, system bandwidth, and latency constraints for the full-duplex UAV-assisted IoT devices. In another work, a single-hop and multi-hop relay-assisted UAV network with a ground-based transmitter and receiver, the problem of finding the optimal UAV location to minimize the impact of interference was investigated \cite{hosseinalipour2020interference}. A theoretical framework was presented to determine the minimum number of UAVs required and their optimal locations to meet the minimum requirement of the received average signal-to-interference ratio. Considering a heterogeneous wireless network of a ground base station, relay aerial vehicles, and a high-altitude platform, the end-to-end delay and reliability analysis of downlink communication links was investigated \cite{salehi2022ultra}. It was demonstrated that even with a very efficient interference mitigation technique employed, the current wireless networks designed for terrestrial users are not suitable to meet the URLLC requirements for downlink control communication to aerial vehicles.

To improve the communication performance and network connectivity for the ground nodes or vehicles in a 3D urban scenario, particle swarm optimization was utilized to find the optimal UAV positions functioning as relay nodes \cite{ladosz2016optimal}. To validate the proposed approach, an indoor experiment was also conducted. However, it is to be noted that, there were no communications among the relayed UAVs. In another work, a model predictive control approach was utilized to construct an energy-efficient communication link between the ground nodes with UAV operating as a relay node \cite{faqir2018energy}. In particular, a problem was formulated to optimize the source’s transmit energy and the UAV’s propulsion energy by jointly optimizing the UAV's mobility and transmission across the multiple access channel. However, the approach was limited to a linear trajectory where it was assumed that the UAV is moving along a single dimension at a fixed height with no lateral displacement. In \cite{zeng2016throughput}, a throughput maximization problem was formulated for a UAV-assisted mobile relay network. The problem was based on jointly optimizing the trajectory of the relay UAV node and the power allocation of the source and relay node, subject to the information-causality constraints. In particular, the authors considered two scenarios. In the first scenario, the relay trajectory was kept fixed, and it was shown that the optimal source and UAV power allocations obey a staircase water-filling structure with non-increasing and non-decreasing power levels at the source and UAV, respectively. Whereas, in another scenario, an iterative algorithm was presented to jointly optimize the UAV trajectory and the power allocation in an alternating way. In another similar work \cite{ono2016wireless}, however, considering the circling operation of the fixed-wing UAV, a rate optimization approach is presented for UAV enabled wireless relay network. It was assumed that there is no direct link between the ground-based source and the destination. A variable rate protocol was developed to optimally adjust the data rate depending on the location of the UAV. It was shown that the proposed approach significantly outperforms the conventional fixed-rate relaying network.

Since there is no regulatory and well-defined pre-allocated spectrum band for UAV communications, it makes constructing a UAV-assisted communication network a non-trivial task. Therefore, UAV-assisted network generally coexists with other wireless networks, e.g., cellular networks \cite{rahmati2019energy}. Thus, formulating the problem of delay minimization given the reliability constraint is critical. This is the main fact that motivated us in formulating the information-theoretic-based framework to investigate the end-to-end transmission delay in the presence of interference. Utilizing our existing outdoor 5G testbed \cite{Wang2021DevelopmentResearch}, Over the Air (OTA) verification and extension of the proposed work is being performed. 

\subsection{Contributions}
Unlike traditional terrestrial or satellite communication networks, the air-to-ground integrated network is affected by the limitations arising simultaneously from the following segments, i.e., from the aspects of mobility management, power control, and end-to-end QoS requirements. Therefore, given the practical resource constraints of an air-to-ground integrated network, it is critically important for an integrated network to achieve optimal and reliable performance given the restriction in power consumption. Therefore, optimal system integration and network design and configuration are of great significance in an air-to-ground and air-to-air integrated network. To this end, we present a novel information-theoretic approach to the design of an optimal air-to-ground integrated network that assesses and minimizes end-to-end transmission latency in the presence of interference, white noise, and potential jamming. The main contributions of this paper are listed below.
\begin{itemize}
    \item We present an information-theoretic framework for the design of an optimal ground-to-UAV communication network that minimizes end-to-end transmission delay. The proposed framework is useful as it describes the minimum transmission latency a UAV network must satisfy while achieving a given level of reliability in terms of the average error probability. In particular, we first derive a tight upper bound for the capacity of the air-to-ground integrated channel in the presence of interference and white noise. Subsequently, given the reliability constraint, a framework is developed to analyze the delay introduced in the channel.
    \item Further, considering the general air-to-ground network with inter-relay communication, the results are presented to comprehensively characterize the optimal performance of the system towards end-to-end delay minimization. We show that despite the simplicity of the point-to-point direct link, the latency of the amplify-and-forward (AF) relayed-based link can be lower if the relay is properly located.
    \item In addition, we considered the power consumption restriction in the air-to-ground and air-to-air networks. Within the range of allowable system latency, the optimal transmit power is derived based on the location information of a UAV. From the analysis, it is shown that increasing the transmission power is not always the proper solution to the delay minimization problem.
\end{itemize}

The rest of the paper is organized as follows. In Section \ref{SYSTEM_MODEL}, we present the system model. The delay analysis is presented in Section III whereas results and discussions are depicted in Section IV. Finally, the conclusion is drawn in Section V.

     \begin{table}
	\renewcommand{\arraystretch}{1.4}
	\caption{Notations}
	\label{table_example}
	\centering
	\begin{tabular}{ l p{6cm} } 
		\hline
		\textbf{Parameter} & \textbf{Description} \\
		& \\
		\hline
		$ d_c $ & Block length of the code word $s_T$ \\
		\hline
        $R_c$ & Code rate \\
        \hline
        $ P_T $ & Transmit power \\
		\hline
        $P_I^{(i)}$ & Transmit power of the $i$th interfering node \\
        \hline
        $P_N$ & Power of the amplifying node \\
        \hline
        $\phi_e$ & Reliability constraint \\
        \hline
        $\Phi_I$ & Number of active interfering nodes at any time $t$\\
        \hline
        $h_{u,v}$ & channel gain between the nodes $u$ and $v$ \\
        \hline
        $\alpha_{u,v}$ & Path loss exponent of the link $u \rightarrow v$\\
        \hline
        $\alpha_{I,v}^{(i)}$ & Path loss exponent of the link between the $i$th interferer and the $v$th receiver\\
        \hline
        $\theta_{u,v}$ & Elevation angle between the transmitting node $u$ and the receiving node $v$ \\
        \hline
        $ \theta_{I,v}^{(i)}$ & Elevation angle between the receiver $v$ and the $i$th interfering node \\
        \hline
        $d_{u,v}$ & 3D distance between the nodes $u$ and $v$ \\
        \hline
        $w_R(t) $ & AWGN noise at the input of the receiver \\
        \hline
        $w_{N}(t)$ & AWGN noise at the input of the relay node \\
        \hline
        $N_0$ & Noise spectral density \\
        \hline
        $B_{uv}$ & Link bandwidth between the nodes $u$ and $v$ \\
        \hline
        $d_{u,v}$ & 3D distance between the nodes $u$ and $v$ \\
        \hline
        $B$ & Information bits in a codeword message of length $d_c$ \\
        \hline
		\end{tabular}
\end{table}

\section{System Model}\label{SYSTEM_MODEL}

\begin{figure}
	\centering
	\includegraphics[width=0.5\textwidth]{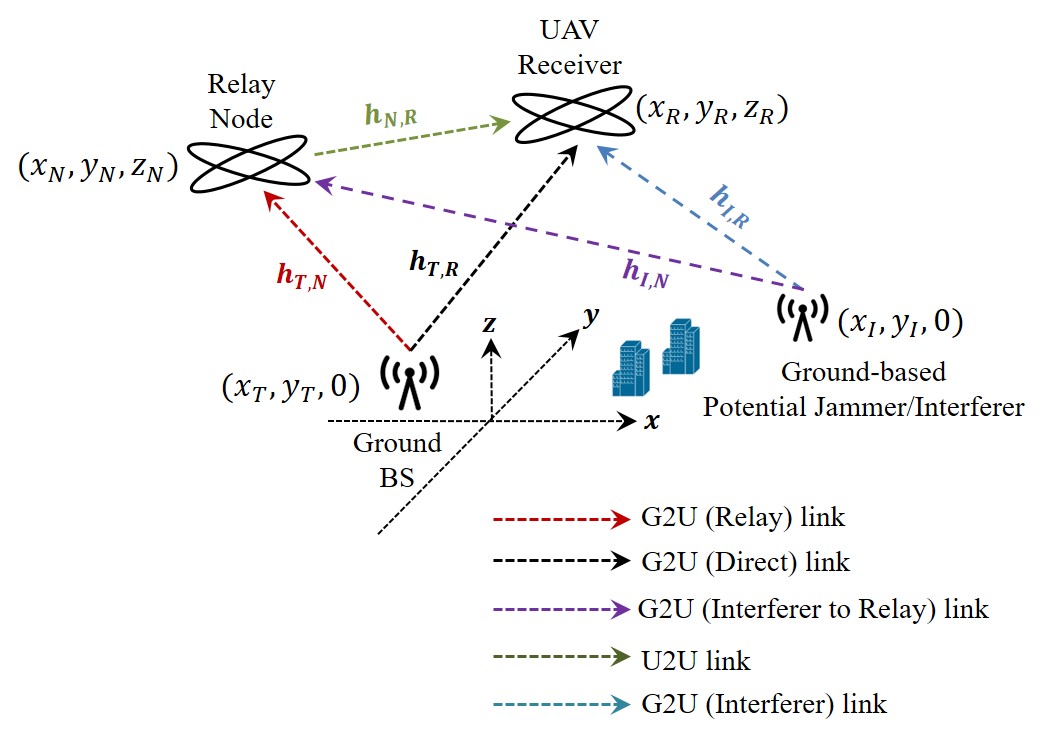}
	\caption{Illustration of the system model.}
	\label{FIG-1}
\end{figure}

We consider an interference network where a transmitted signal is corrupted by the interference signal plus the additive white Gaussian noise (AWGN). In this noise-plus interference-limited network, there can be two types of communication links: a direct link from the ground base station (BS) to the receiver UAV and a relay-assisted link from the BS to the receiver UAV. Let the ground BS, located at $(x_T, y_T, 0)$, communicates to the receiver $(R_x)$ located at $(x_R, y_R, z_R)$ in the presence of interfering nodes located at $(x_I^{(i)}, y_I^{(i)}, 0), i \in \Phi_I$. $\Phi_I$ is a subset containing all interfering nodes transmitting at time $t$.

Let the BS transmits with power $P_T$ over a channel with path loss exponent $\alpha_T$ and fading coefficient $h_T$. $d_T  = \sqrt {\left( {x_T  - x_R } \right)^2  + \left( {y_T  - y_R } \right)^2  + \left( {z_T  - z_R } \right)^2 }$ is the distance between the BS and $R_x$. $s_T$ is a code word of length $N$ and is a sequence of $N$ numbers such that $s_T = (s_{T_1}, s_{T_2}, \cdot\cdot\cdot, s_{T_N}).$ The code word $s_T$ for an underlying ground-to-UAV channel may be thought of geometrically as a point in $N-$dimensional Euclidean space. The impact of the channel impairment due to the Gaussian noise and interference is then to move $s_T$ to a nearby point according to a spherical Gaussian distribution. $s_T(t)$ are random symbols drawn from a constellation size $M$ with unitary mean power. $P_I^{(i)}$ denotes the transmit power of the $i$th interfering node.

We consider an AF relay channel with one hop, as illustrated in Fig. \ref{FIG-1}. To avoid interference between the ground BS and the relay, it is assumed that the information transmission is conducted via time division where the transmission from the ground BS to the receiver is divided into two time slots. In the first slot, the signal is received by the relaying UAV, which is then amplified, and in the second slot, the amplified signal is received by the receiver UAV.  An arbitrary relay UAV can normalize and re-transmit the received signal. With the power $P_N$, the re-transmitted signal $s_N$ is represented by
\begin{equation}
\begin{split}
  &  s_N(t)  = \frac{{\sqrt {P_N } Y_N \left( t \right)}}{{\sqrt {E\left[ {\left| {Y_N \left( t \right)} \right|^2 } \right]} }} \\ 
  &= \frac{{\sqrt {P_N } \left( {\sqrt {P_T d_{T,N}^{ - \alpha _{T,N} \left( {\theta _{T,N} } \right)} } h_{T,N} s_T \left( t \right) + w_N \left( t \right)} \right)}}{{\sqrt {P_T d_{T,N}^{ - \alpha _{T,N} \left( {\theta _{T,N} } \right)} \left| {h_{T,N} } \right|^2  + B_{TN} N_0 } }} \\
\end{split}
\label{EQ-1},
 \end{equation}
 where $P_T$ is the transmit power of the ground BS and $w_N$ denotes the white noise at the input of the relaying UAV. $E[\cdot]$ represents the expectation operation. Following (\ref{EQ-2}), the signal received at the receiver is expressed as
 \begin{equation}
     \begin{split}
& Y_R\left( t \right) = s_N(t) \sqrt {d_{N,R}^{ - \alpha _{N,R} \left( {\theta _{N,R} } \right)} } h_{N,R}  \\ 
 & + \sum\limits_{i \in \Phi _I } {\sqrt {P_I^{\left( i \right)} d_{i,R}^{ - \alpha _I^{\left( i \right)} \left( {\theta _I^{\left( i \right)} } \right)} } h_{I,R}^{\left( i \right)} s_I^{\left( i \right)} \left( t \right)}  + w_R \left( t \right) \\
     \end{split}
     \label{EQ-2}
 \end{equation}

We consider that the G2A and the A2A communication links experience LOS propagation with LOS probability $P_{LOS,h}$, $h \in \{h_T, h_{T, N}, h_{N, R}, h_{I, R}, h_{I, N}\}$, expressed as \cite{series2013propagation}
\begin{equation}
P_{LOS,h}  = \frac{1}{{1 + f_1 \exp \left( { - f_2 \left( {\theta _h  - f_1 } \right)} \right)}}
\label{EQ-3},
\end{equation}
where $f_1$ and $f_2$ are environment-dependent parameters determined by the building density and heights.
Due to the LOS path for all the described links, the small-scale channel fading gain $h$ can be assumed to follow the Rician model with PDF given by
\begin{equation}
f_h \left( h \right) = \frac{h}{{\sigma _h^2 }}\exp \left( { - \frac{{h^2  + \rho _h^2 }}{{2\sigma _h^2 }}} \right)I_0 \left( {\frac{{h\rho _h }}{{\sigma _h^2 }}} \right), h \geq 0
\label{EQ-4},
\end{equation}
where $\sigma_h$ and $\rho_h$ represent the strength of the LOS and NLOS components. $I_0(\cdot)$ is the zeroth order modified Bessel function of the first kind. Following (\ref{EQ-4}), the Rice factor $K_h$(dB) of any link $h \in \{h_{T,R}, h_{T, N}, h_{N, R}, h_{I, R}, h_{I, N}\}$, can then be defined as $K_h$(dB) $= 10\log_{10}(\frac{\rho_h^2}{2\sigma_h^2})$. We like to point out that the Rice factor $K_h$ is a function of the parameters such as the carrier frequency and the elevation angle $\theta_h $ \cite{salehi2022ultra}.

We define the instantaneous signal-to-noise ratio (SNR) of arbitrary link $u \rightarrow v$, from the transmitter $u, u \in \{T, N\}$ to the receiver $v, v \in \{N, R\}$,  over the known interference channel as
\begin{equation}
\lambda_{uv}  = \frac{{\left| {\sqrt {P_u d_{u,v}^{ - \alpha _{uv} \left( {\theta _{uv} } \right)} } h_{u,v} } \right|^2 }}{{B_{uv}N_0 }}
\label{EQ-5},
\end{equation}
and the power ratio between the intended received signal at $v, v \in \{N, R\}$ from $u, u \in \{T, N\}$, and the interference signal from the $i$th interfering node is expressed as
\begin{equation}
\gamma_{uv}^{(i)}  = \frac{{\left| {\sqrt {P_u d_{u,v}^{ - \alpha _{u,v} \left( {\theta _{u,v} } \right)} } h_{u,v} } \right|^2 }}{{\left| {\sum\limits_{i \in \Phi _I } {\sqrt {P_I^{\left( i \right)} d_{i,v}^{ - \alpha _{I,v}^{\left( i \right)} \left( {\theta _{I,v}^{\left( i \right)} } \right)} } h_{I,v}^{\left( i \right)} } } \right|^2 }}.
\label{EQ-6}
\end{equation}
The expression in (\ref{EQ-6}) is general enough to be treated as a signal-to-interference ratio (SIR). 

Averaging over the fading statistics described in (\ref{EQ-4}), we obtain the average received SNR $\bar{\lambda}_{uv}$ as illustrated in (\ref{EQ-7}). $G_{m,n}^{s,t}[\cdot]$ in (\ref{EQ-7}) represents the Meijer G-function \cite{gradshteyn2014table}. The derivation steps are provided in Appendix \ref{APPENDIX-1}.

\begin{figure*}
    \begin{equation}
\bar \lambda _{uv}  = \frac{{\sqrt {P_u d_{u,v}^{ - \alpha _{u,v} \left( {\theta _{u,v} } \right)} } \left( { - 2\pi \sigma _{h_{u,v} }^2 } \right)}}{{B_{uv} N_0 }}\exp \left( { - \frac{{\rho _{h_{u,v} }^2 }}{{2\sigma _{h_{u,v} }^2 }}} \right)G_{3,4}^{2,1} \left[ {\frac{{2\rho _{h_{u,v} }^2 }}{{\sigma _{h_{u,v} }^2 }}\left| {\begin{array}{*{20}c}
   { - 2, - 1,\frac{1}{2}}  \\
   {0, - 2,0,\frac{1}{2}}  \\
\end{array}} \right.} \right]
    \label{EQ-7}.
    \end{equation}
    \begin{center}
    \line(1,0){380}
 \end{center}
\end{figure*}

\section{Delay Analysis}
The critical and important questions we would like to answer in this work are: What is the transmission latency of the G2U communication link under the given system parameters$?$ And what are the optimal locations of the relay and the UAV to have a positive impact in minimizing the transmission delay$?$ More precisely, our objective is to minimize the delay in transmitting messages between the transmitter and the receiver in UAV communication while guaranteeing a required level of reliability $\phi_e$, such that, 
\begin{equation}
P_e \left( {d_c,R_c} \right) \le \phi _e,
\label{EQ-8}
\end{equation}
where $P_e \left( {d_c,R_c} \right)$ represents the average error probability of the code with length $d_c$ and rate $R_c$. The channel code rate $R$ is a function of the input target signal distribution $f_T$, the input interference signal distribution $f_I$, $\lambda$, and $\gamma$. The channel code rate $R_c$ is achievable if there is an encoding function to map each transmitted message $S$ to $s_T$ and there is another decoding function to map each received signal $Y$ and interference information to a transmitted message as $\hat{S}$, such that the average error probability $P_e(d_c, R_c) \triangleq P_r[S \ne \hat{S}] \rightarrow 0$ when $N$ goes large. Following this, we define the reliability function in terms of the error exponent of the channel as \cite{gallager1968information}
\begin{equation}
E\left( R_c \right) =  - \mathop {\lim }\limits_{d_c \to \infty } \sup \frac{{\ln P_{e,\min } \left( {d_c,R_c} \right)}}{d_c},
\label{EQ-9}
\end{equation}
where ${\rm sup}(\cdot)$ represents the supermum function and $P_{e,\min }\left( {d_c, R_c} \right)$ denotes the infimum or the greatest lower bound of the error probability over all $\left( {d_c, R_c} \right)$ codes for a given $d_c$ and $R_c$. It can be readily shown that the error exponent depicted in (\ref{EQ-3}) is upper bounded by the sphere packing exponent
\begin{equation}
E\left( R_c \right) \ge \mathop {\max }\limits_{\rho  > 0} \left\{ {E_0 \left( \rho  \right) - \rho R_c} \right\},
\label{EQ-10}
\end{equation}
where
\begin{equation}
E_0 \left( \rho  \right) =  - \log \int\limits_{ - \infty }^\infty  {\left( {\int {\left( {P\left( {Y_R\left| s_T \right.} \right)} \right)^{\frac{1}{{1 + \rho }}} dP\left( s_T \right)} } \right)^{{{1 + \rho }}} } dY_R
\label{EQ-11}
\end{equation}
where $P\left(s_T\right)$ denotes the cumulative distribution function of the input $s_T$ and $P\left( {Y_R\left| s_T \right.} \right)$ is the probability density (or mass) function of the output $Y$ given the input $s_T$. $\rho $ is a parameter, $0 \leq \rho \leq 1$, and should be selected such that $d_c$ is minimized. Utilizing (\ref{EQ-7}), (\ref{EQ-8}), and (\ref{EQ-10}), and applying the identity \cite[eq. 11]{adamchik1990algorithm}, the minimum transmission delay introduced into the channel in order to satisfy the reliability constraint is given by
\begin{equation}
d_c  \ge \frac{{\rho B - \log \phi _e }}{{G_{2,2}^{1,2} \left[ {\bar \lambda _{uv} \left| {\begin{array}{*{20}c}
   {1,1}  \\
   {1,0}  \\
\end{array}} \right.} \right]}}
\label{EQ-12A},
\end{equation}
where $B$ represents the information bits in the codeword of length $d_c$.

We define the capacity of the uplink ground-to-UAV channel impaired by the known interference as
\begin{equation}
\begin{split}
   & C\left( {\lambda ,\gamma } \right) = \mathop {\sup }\limits_{f_T } \left[ {\mathop {\inf }\limits_{f_I } R_c\left( {f_T ,f_I ,\lambda ,\gamma } \right)} \right] \\ 
 & \hspace{0.5cm}{\rm subject \hspace{0.1cm} to } \hspace{0.1cm}\mathbb{E}\left[ {s_I } \right]^2  = 1, \mathbb{E}\left[ {s_T } \right]^2  = 1 \\
\end{split}
\label{EQ-12}
\end{equation}
Next, we present a tight upper bound for $C\left( {\lambda ,\gamma } \right)$ defined in (\ref{EQ-12}).

To decode the transmitted message $S$ correctly with a low probability of error, the conditional entropy $H(S|Y_R, s_I)$ has to be close to zero \cite{tse2005fundamentals}. Following this, and applying Fano's inequality \cite{scarlett2019introductory} into the definition of channel entropy, it can be readily shown that
\begin{equation}
d_cR_c\left( {f_T ,f_I ,\lambda ,\gamma } \right) \le I\left( {S;Y_R,s_I } \right) + d_c\zeta _{d_c},
\label{EQ-9}
\end{equation}
where $\zeta_N$ is the error detection parameter satisfying the condition $\mathop {\lim }\limits_{d_c \to \infty } \zeta _{d_c}  = 0$. Applying the definition of mutual information and utilizing the fact that $S$ and $s_I$ are independent, (\ref{EQ-9}) can be written as
\begin{equation}
d_cR_c\left( {f_T ,f_I ,\lambda ,\gamma } \right) \le H\left( {Y_R\left| {s_I } \right.} \right) - H\left( {Y_R\left| {S,s_I } \right.} \right) + \zeta _{d_c}
\label{EQ-14},
\end{equation}
where $H\left( {Y_R\left| {s_I } \right.}\right)$ can be interpreted as the uncertainty in $Y$ conditional on $s_I$. $H\left( {Y_R\left| {S,s_I } \right.} \right)$ accounts for the reduction in uncertainty of $Y$ conditional on $s_I$ from the observation of $s_I$. Substituting the expressions for $H\left( {Y_R\left| {s_I } \right.}\right)$ and 
$H\left( {Y\left| {S,s_I } \right.} \right)$ into (\ref{EQ-14}), the tight upper bound for the capacity of the air-to-ground channel is derived as illustrated in (\ref{EQ-15}). $N_p$ in the summation term in (\ref{EQ-15}) denotes the packet length. $T_c$ represents the block of symbols over which the channel is assumed to be constant. The detailed derivation steps of (\ref{EQ-15}) are provided in Appendix \ref{APPENDIX-2}.

\begin{figure*}[t]
    \begin{equation}
    \begin{split}
        & R_{c,uv} \le \frac{{\left( {T_c - 1} \right)}}{T_c}\log \left( {1 + \frac{{P_u d_{u,v}^{ - \alpha _{u,v} \left( {\theta _{u,v} } \right)} \left\| {h_{u,v} } \right\|^2 }}{{B_{uv}N_0 }}} \right) \\
        & \hspace{0.4cm}+ \frac{1}{N_p}\sum\limits_{j = 1}^{N_p/T_c} {\mathop E\limits_{s_{I,j} } \left[ {\log \left( {1 + \frac{{P_u d_{u,v}^{ - \alpha _{u,v} \left( {\theta _{u,v} } \right)} \left\| {h_{u,v} } \right\|^2 }}{{\sum\limits_{i \in \Phi _I } {P_I^{\left( i \right)} d_i^{ - \alpha _{I,v}^{\left( i \right)} \left( {\theta _{I,v}^{\left( i \right)} } \right)} \left\| {h_{I,v}^{\left( i \right)} } \right\|^2 \left\| {s_{I,j}^{\left( i \right)} } \right\|^2  + {B_{uv}N_0 } } }}} \right)} \right]}  + \zeta _{d_c}
    \end{split}
\label{EQ-15}.
    \end{equation}
    \begin{center}
    \line(1,0){380}
 \end{center}
\end{figure*}

\begin{figure*}
    \begin{equation}
P_{u,{\rm min}}  \ge \left\{ {\left( {1 + \rho } \right)\exp \left( {\frac{{\rho B - \log \phi _e }}{{d_{c,\max } }}} \right)\left( {\left| {\sum\limits_{i \in \Phi _I } {\sqrt {P_I^{\left( i \right)} d_{I,v}^{ - \alpha _{I,v}^{\left( i \right)} \left( {\theta _{I,v}^{\left( i \right)} } \right)} } h_{I,v}^{\left( i \right)} } } \right|^2  + B_{uv}N_0 } \right) - 1} \right\}\frac{1}{{d_{u,v}^{ - \alpha _{u,v} \left( {\theta _{u,v} } \right)} \left\| {h_{u,v} } \right\|^2 }}
    \label{EQ-16}.
    \end{equation}
    \begin{center}
    \line(1,0){380}
 \end{center}
\end{figure*}

\begin{figure*}
    \begin{equation}
        \begin{split}
 \bar \lambda _{u,v}  = \frac{{\sqrt {P_x d_{u,v}^{ - \alpha _{u,v} \left( {\theta _{u,v} } \right)} } }}{{B_{uv} N_0 }}\left( {\frac{\pi }{{\sigma _{h_{u,v} }^2 }}} \right)\exp \left( { - \frac{{\rho _{h_{u,v} }^2 }}{{2\sigma _{h_{u,v} }^2 }}} \right)\left\{ {\int\limits_0^\infty  {\left\{ {\underbrace {h_{u,v}^3 G_{1,3}^{1,0} \left[ {h_{u,v}^2 \frac{{\rho _{h_{u,v} }^2 }}{{\sigma _{h_{u,v} }^2 }}\left| {\begin{array}{*{20}c}
   {\frac{1}{2}}  \\
   {0,0,\frac{1}{2}}  \\
\end{array}} \right.} \right]}_{{\rm Term \hspace{0.15  cm} I}}} \right.} } \right. \\ 
 \left. { - \left. {\underbrace {h_{u,v}^3 G_{1,2}^{1,1} \left[ {\frac{{h_{u,v}^2 }}{{2\sigma _{h_{u,v} }^2 }}\left| {\begin{array}{*{20}c}
   1  \\
   {1,0}  \\
\end{array}} \right.} \right]G_{1,3}^{1,0} \left[ {h_{xy}^2 \frac{{\rho _{h_{u,v} }^2 }}{{\sigma _{h_{u,v} }^2 }}\left| {\begin{array}{*{20}c}
   {\frac{1}{2}}  \\
   {0,0,\frac{1}{2}}  \\
\end{array}} \right.} \right]}_{{\rm Term \hspace{0.15 cm}II}}} \right\}dh_{u,v} } \right\} \\
        \end{split}
        \label{EQ-A1}.
    \end{equation}
    \begin{center}
    \line(1,0){380}
 \end{center}
\end{figure*}

\begin{figure}
	\centering
	\includegraphics[width=0.47\textwidth]{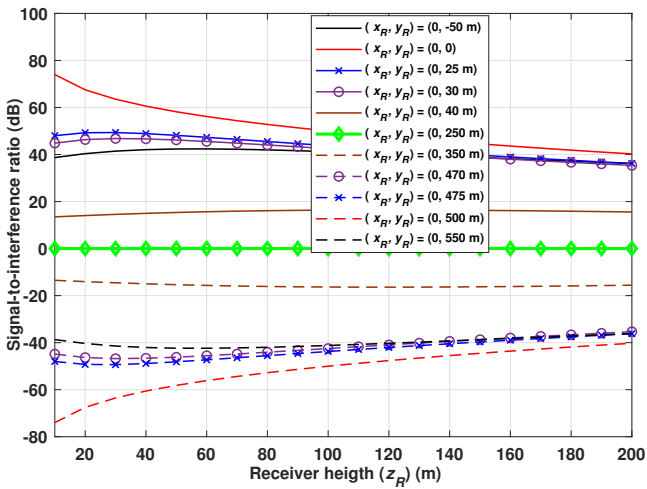}
	\caption{Signal-to-interference ratio relative to the receiver height for different $y_R$ along the line joining the transmitter and the interferer.}
	\label{FIG-2}
\end{figure}

We like to point out that, since the channel is random over a single transmission interval with the assumption that the channel state information is not available to the BS, and considering the capacity as defined in (\ref{EQ-12}), it is now possible that $C\left( {\lambda, \gamma } \right) = 0$ with nonzero probability irrespective of the link range and transmission power. Therefore, the power allocation approach may not be an optimal solution for the uplink G2U networks. Moreover, it is now impossible to guarantee successful information transmission with $\phi_e = 0$ and $C\left( {\lambda, \gamma } \right) > 0$.

Next, we obtain the minimum transmit power $P_{u, \rm min}$ required at node $u$, given the maximum allowable delay $d_{c, \rm max}$ and the reliability constraint $\phi_e$ at the receiver $v$. Following (\ref{EQ-2}), (\ref{EQ-10}), and (\ref{EQ-11}), the tight upper bound on the transmit power requirement, subject to the reliability constraint and the maximum delay threshold, can readily be derived as shown in (\ref{EQ-16}).


\section{Results and Analysis}
The simulation parameters are summarized in Table \ref{TABLE-II}. We consider a slow-fading channel, where the actual channel gains are assumed to be constant, however, random over a single transmission interval. Unless otherwise stated, the interferer coordinates are set to $(0, 500, 0)$. To get a satisfactory statistical average, for each link configuration parameter set, 1000 realizations of normal distributed independent random  variables are generated. The delay is then obtained by averaging the results over 1000 realizations. For brevity, we assume one interfering node. For simplicity, we consider the following three cases to describe the link geometries.

\begin{table}
	\renewcommand{\arraystretch}{1.4}
	\caption{System Parameters}
	\label{TABLE-II}
	\centering
	\begin{tabular}{ l p{2.5cm} } 
		\hline
		\textbf{Parameter} & \textbf{Value} \\
		& \\
		\hline
		Packet size & $32$ bytes \\
		\hline
            Rice factor of G2A link & $5 \sim 12$ dB \\
		\hline
            Rice factor of A2A link & $ 10 \sim 12 $ dB \\
		\hline
            Noise spectral density $N_0$ \cite{salehi2022ultra} & $- 174$ dBm/Hz \\
            \hline
            LOS (NLOS) shadow fading standard deviation & $4$ $ (6)$ dB \\
            \hline
            Model parameter $f_1$ \cite{kim2018outage} &    12.08 \\
            \hline
            Model parameter $f_2$ \cite{kim2018outage} &    0.11   \\
            \hline
            Carrier frequency $f_c$ & $2$ GHz \\
            \hline
        \end{tabular} 
\end{table}

\begin{itemize}
    \item {\bf{Case I:}} It considers a scenario where the receiver is located above in a region near the vicinity of the transmitter
    \item {\bf{Case II:}} It considers a scenario where the receiver is located above in a region near the midway of the line joining the transmitter and the receiver
    \item {\bf{Case III:}} It considers a scenario where the receiver is located above in a region near the vicinity of the interfering node
\end{itemize}

Fig. \ref{FIG-2} shows the SIR variation with receiver height $(z_R)$ at different positions along the line joining the BS and the interferer. For Case I, as $z_R$ increases, the signal of interest becomes weaker, and hence, the interference becomes relatively stronger. On the contrary, for Case III, increasing $z_R$ results in a weaker interference thereby making the signal of interest relatively stronger and hence higher SIR. However, the nearly constant SIR for case II can be attributed to the fact that both, the signal of interest and the interference signal, become weak proportionally with increasing $z_R$.

\begin{figure}
    \centering
    \begin{subfigure}[b]{0.46\textwidth}
        \centering
        \includegraphics[height=2.4in]{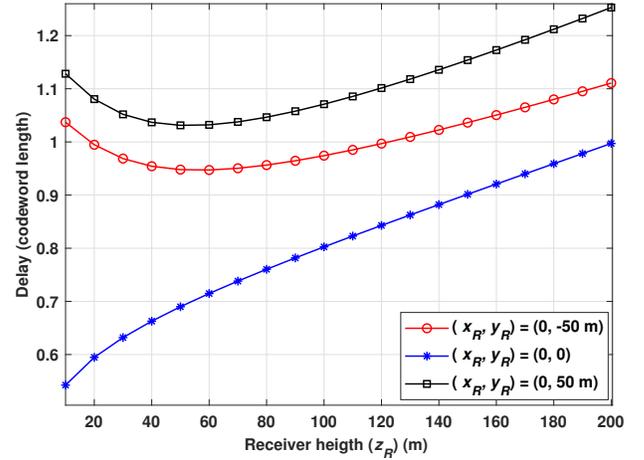}
        \caption{Interference-limited channel.}
        \label{FIG-3A}
    \end{subfigure}%
   
    \begin{subfigure}[b]{0.46\textwidth}
        \centering
        \includegraphics[height=2.4in]{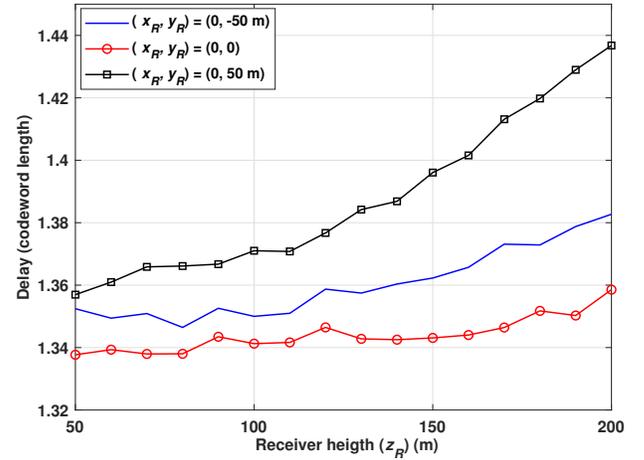}
        \caption{Interference channel with end-to-end received 
signal-to-noise power ratio $\lambda = 30$ dB.}
    \label{FIG-3B}
    \end{subfigure}

    \begin{subfigure}[b]{0.46\textwidth}
        \centering
        \includegraphics[height=2.4in]{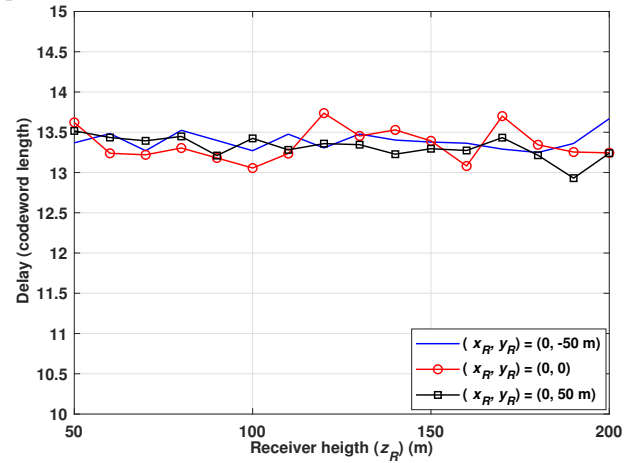}
        \caption{Interference channel with end-to-end received 
signal-to-noise power ratio $\lambda = 0$ dB.}
    \label{FIG-3C}
    \end{subfigure}

    \caption{Delay profile as a function of the receiver height over different channel conditions (Case I). }
    \label{FIG-3}
\end{figure}

In Fig. \ref{FIG-3A}, \ref{FIG-3B}, and \ref{FIG-3C}, we show the results for the delay, while varying the receiver height $z_R$ for several channel conditions, i.e., over the interference-limited channel, and when $\bar{\lambda} = 30$ dB and $\bar{\lambda} = 0$ dB, respectively. In Fig. \ref{FIG-3A}, as expected, we can observe that for an interference-limited channel, when the receiver is located above the transmitter, the delay is an increasing function of the receiver height. In contrast, when the receiver moves in either direction in a line joining the transmitter and the interferer, the delay shows non-monotonic behaviors with the receiver height. Indeed, for the given link configuration and the placements of the ground BS and the interferer, the delay first decreases with increasing the height till the optimal height is obtained and then increases with the height. This quasi-monotonic behavior can be attributed to the fact that increasing the receiver height beyond the optimal height causes higher interference and path loss. In contrast to the interference-limited channel, slightly different trends are observed in the interference-plus-noisy channel. For example, in Fig. \ref{FIG-3B}, when $\bar{\lambda} = 30$ dB, the delay increases with $z_R$ irrespective of $x_R$ and $y_R$, however, in Fig. \ref{FIG-3C}, when $\bar{\lambda}$ is very low, the three curves, however, cannot be distinguished. It shows that noise is a dominating factor in this case.

\begin{figure}
    \centering
    \begin{subfigure}[b]{0.46\textwidth}
        \centering
        \includegraphics[height=2.4in]{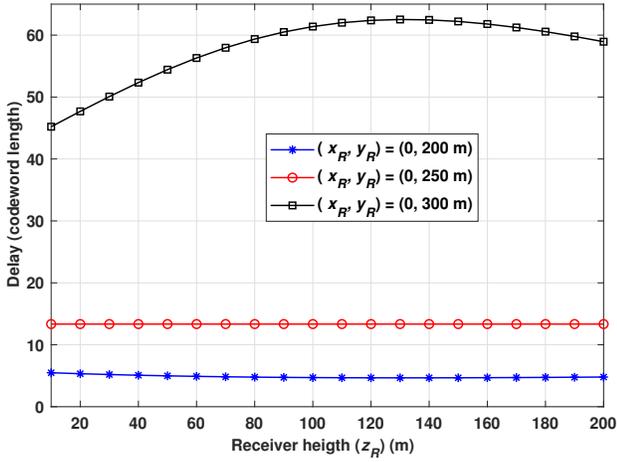}
        \caption{Interference-limited channel.}
        \label{FIG-4A}
    \end{subfigure}%
   
    \begin{subfigure}[b]{0.46\textwidth}
        \centering
        \includegraphics[height=2.4in]{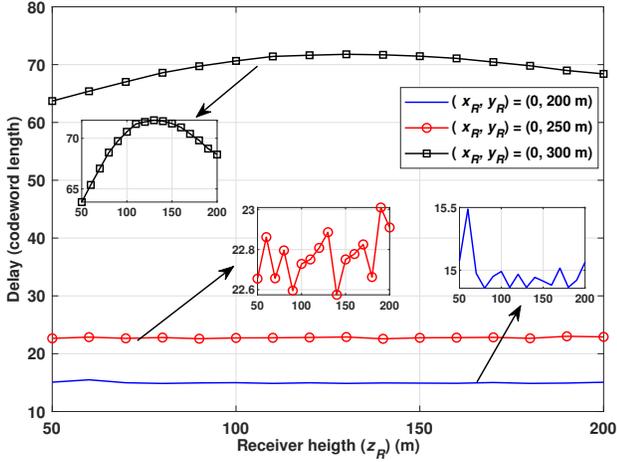}
        \caption{Interference channel with end-to-end received 
signal-to-noise power ratio $\lambda = 0$ dB.}
        \label{FIG-4B}
    \end{subfigure}
    \caption{Delay profile as a function of the receiver height over different channel conditions (Case II). }
    \label{FIG-4}
\end{figure}

Figure \ref{FIG-4} presents the end-to-end delay profile relative to the receiver height for the direct link over different channel conditions for Case II. The different curves in Fig. \ref{FIG-4A} and \ref{FIG-4B} demonstrate the impact of varying the receiver along the mid-point on the line joining the transmitter and the interferer. Based on the results, a few important observations can be made.
\begin{itemize}
    \item If the UAV is located near the midpoint such that, $d_{T, N} > d_{i,R}$, the delay first increases rapidly with the height and then decreases gradually with further increasing the UAV height. This monotonic behavior can be attributed to the fact that increasing the height initially causes received signal power to decrease relatively at a higher rate than interference and thereby reduces the number of packets successfully received. However, at larger heights, the impact of the interference signal also reduces, thus causing the delay to decrease gradually.
    \item Interestingly, and contrary to the above point, if the UAV is located near the midpoint such that $d_{T,N} \leq d_{i,R}$, increasing the height does not reflect significant changes in the delay.
\end{itemize}

\begin{figure}
    \centering
    \begin{subfigure}[b]{0.45\textwidth}
        \centering
        \includegraphics[height=2.4in]{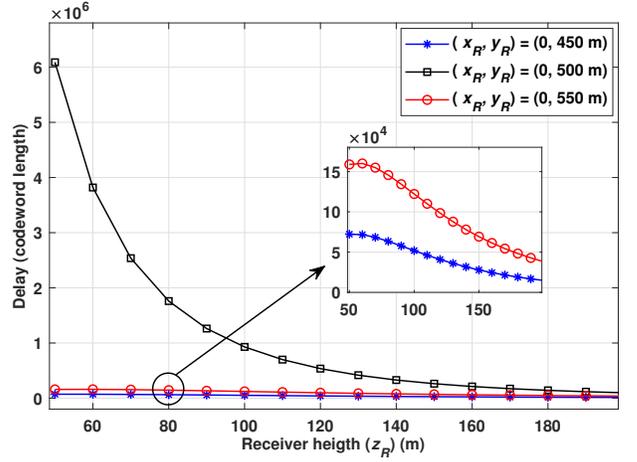}
        \caption{Interference-limited channel.}
    \end{subfigure}%
   
    \begin{subfigure}[b]{0.45\textwidth}
        \centering
        \includegraphics[height=2.4in]{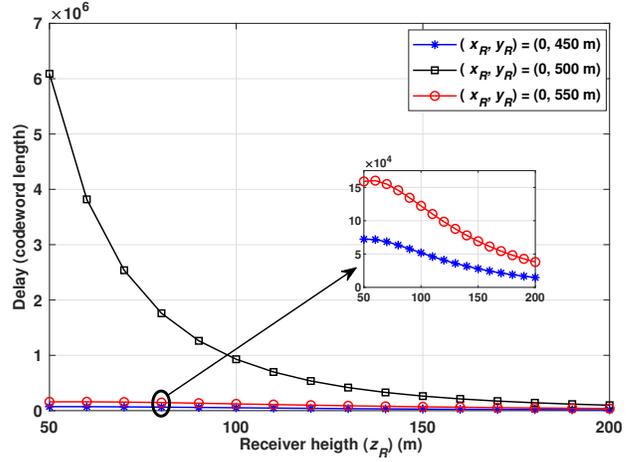}
        \caption{Interference channel with end-to-end received 
signal-to-noise power ratio $\lambda = 0$ dB.}
    \end{subfigure}
    \caption{Delay profile as a function of the receiver height over different channel conditions (Case III). }
    \label{FIG-5}
\end{figure}

Figure \ref{FIG-5} depicts the delay performance relative to the receiver heights over different channel types for Case III. The results reveal important observations on the delay characteristics, that when the receiver is located in a region near the vicinity of the interfering node, the performance is primarily dominated by interference. Interestingly, as can be seen, the delay decreases with the increasing receiver height for all the curves under all channel conditions. Where the delay values are finite but very high. The higher delay is due to the increasing interference that reduces the success probability of transmission. Moreover, it is to be noted that, irrespective of the channel conditions, the performance is primarily dominated by interference and not by the Gaussian noise.

\begin{figure}
	\centering
	\includegraphics[width=0.47\textwidth]{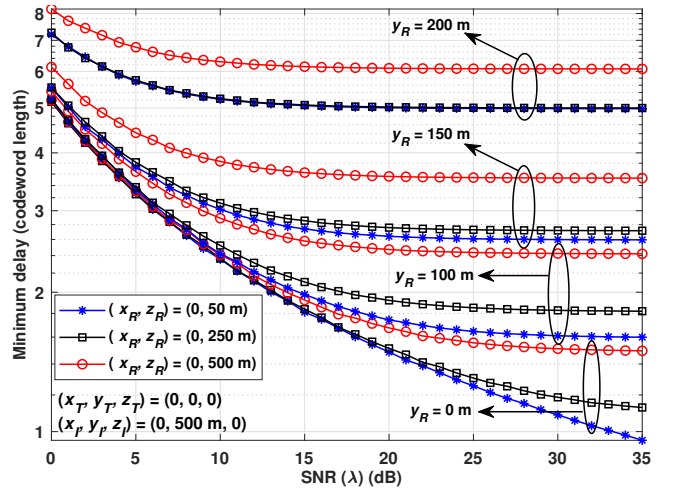}
	\caption{Minimum delay relative to received SNR for different receiver locations.}
	\label{FIG-6}
\end{figure}

Given the location of the receiver and the interferer, the minimum delay a G2U communication system must encounter for different values of $\bar{\lambda}$ is illustrated in Fig. \ref{FIG-6}. In obtaining these curves, the interference power is kept constant at $30$ dBm whereas, the reliability constraint is set to $10^{-4}$. Following important observations can be made.
\begin{itemize}
    \item As can be seen, the impact of increasing the signal power in minimizing the transmission delay dominates only if the UAV is located near a region above the transmitter. However, it is to be noted that the negative rate of change of delay with respect to the change in $\bar{\lambda}$ approaches zero for larger values of $\bar{\lambda}$. Moreover, it is also important to note that, as the UAV approaches a region away from the transmitter and near the interferer, the rate $\frac{dd_{c, \rm min}}{d\bar{\lambda}} \rightarrow 0$ even for smaller values of $\bar{\lambda}$. It shows that increasing $\bar{\lambda}$ gives a negligible gain in terms of minimizing the transmission delay, though may greatly enhance the outage performance.
    \item As low $\bar{\lambda}$ values, increasing the height of the UAV does not necessarily impact the delay performance. However, as $\bar{\lambda}$ increases, the delay performance degrades with increasing the height.
\end{itemize}

\begin{figure}
	\centering
	\includegraphics[width=0.47\textwidth]{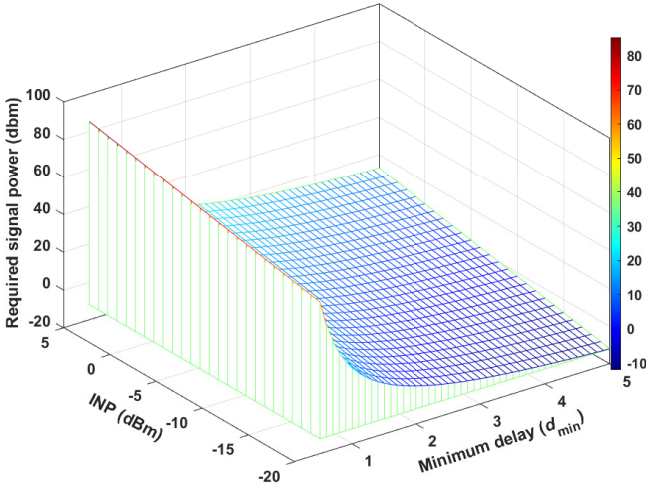}
	\caption{Delay profile relative to the relay node location.}
	\label{FIG-7}
\end{figure}

Given the range of allowable system latency that guarantees that the reliability constraint is met, the minimum transmission power required at the BS for different noise-to-interference power (NIP) levels is illustrated in Fig. \ref{FIG-7}. In obtaining these results, we fix the location of the receiver and the interferer at $(0, 250, 250)$ and $(0, 500, 0)$, respectively. The reliability constraint is set to $P_e \leq \phi_e = 10^{-4}$. The key observations from the results obtained are as follows:
\begin{itemize}
    \item Given the reliability constraint and fixed NIP value, there exists a non-linear relation between the allowable transmission delay and the corresponding transmit power requirements. As can be seen, the required transmit power reduces significantly when the allowable transmission delay is relaxed from its initial value. However, on further relaxing the delay requirement, the required transmit power reduces gradually. This phenomenon validates our claim that increasing the signal power beyond a certain limit may not be an optimal solution to the delay minimization problem.
    \item Given the delay requirement and the reliability constraint, the required signal power increases with increasing NIP. Importantly, it is to be noted that, this behavior follows a constant rate of change of required transmit power with respect to the change in NIP, irrespective of the delay requirement.
\end{itemize}

\begin{figure}
    \centering
    \begin{subfigure}[b]{0.45\textwidth}
        \centering
        \includegraphics[height=2.4in]{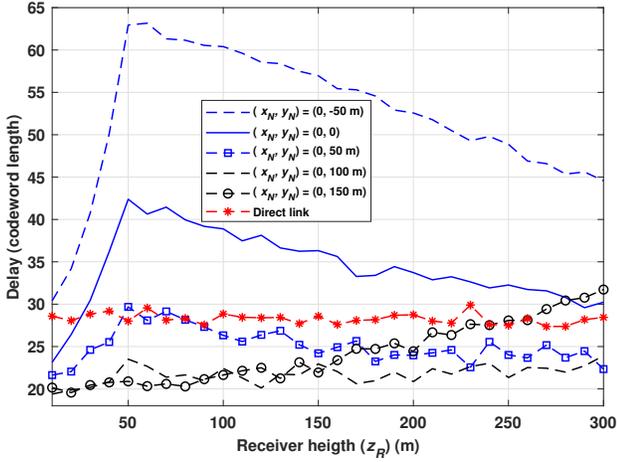}
        \caption{Received signal-to-noise power ratio $\lambda = -2$ dB.}
        \label{FIG-8A}
    \end{subfigure}%
    
    \begin{subfigure}[b]{0.45\textwidth}
        \centering
        \includegraphics[height=2.4in]{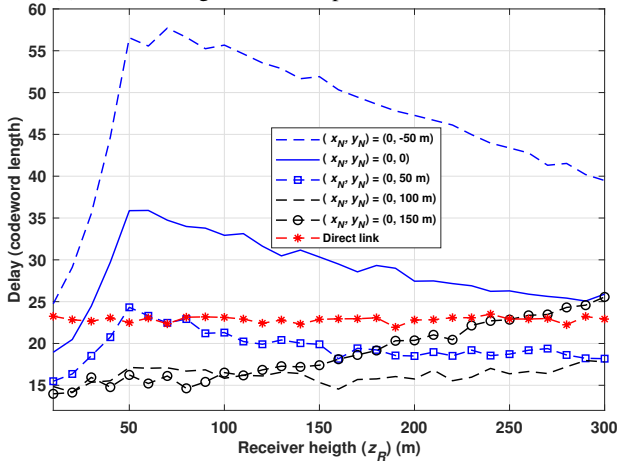}
        \caption{Received signal-to-noise power ratio $\lambda = 0$ dB.}
        \label{FIG-8B}
    \end{subfigure}
    
    \begin{subfigure}[b]{0.45\textwidth}
        \centering
        \includegraphics[height=2.4in]{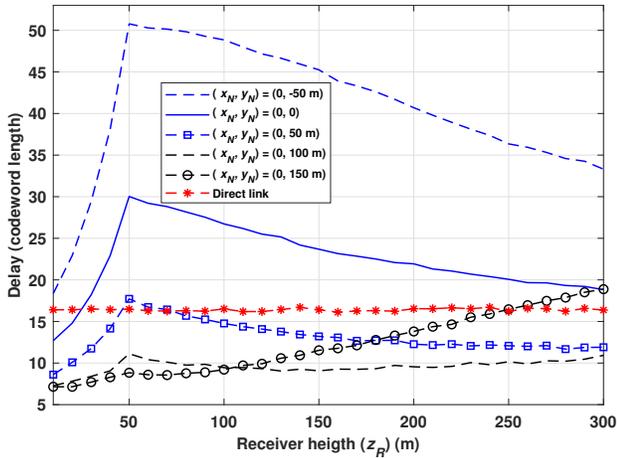}
        \caption{Received signal-to-noise power ratio $\lambda = 5$ dB.}
        \label{FIG-8C}
    \end{subfigure}
    \caption{Delay profile relative to the relay node location for different receiver heights (without noise amplification at the relay node).}
    \label{FIG-8}
\end{figure}

Figure \ref{FIG-8} shows the delay profile against the receiver height for different relay node locations. The $x_R$ and $y_R$ coordinates of the receiver are fixed to $(0, 250)$ and are located midway of the line joining the transmitter and the interferer. The amplification gain of the relay UAV is set to $-3$ dB. However, it is assumed that there is no noise at the input of the relay. For the relay-assisted network, it can be seen that the delay increases with the height initially and then decreases. This phenomenon is more dominating when the relay node is located on the opposite side of the line joining the transmitter and the receiver. Moreover, it can be seen that, as the relay node moves closer to the region between the transmitter and the receiver, the delay performance improves. However, it is to be noted that, as the relay node moves closer to the receiver, the delay though may be lower, but increases with increasing the receiver height. Moreover, for the direct link, as the receiver height increases, there is no impact on the delay performance. Intuitively, this behavior can be attributed to the fact that, if the UAV is located near the region in the middle of the BS and the interferer, changing the UAV height vary the signals strengths from the transmitter and the interferer identically. In obtaining the results in Figs. \ref{FIG-8A}, \ref{FIG-8B}, \ref{FIG-8C}, a few points are important to note.
\begin{itemize}
    \item For the relayed-assisted network, we assume that there is no direct path between the BS and the receiver UAV. Thus, the relayed UAV in our case serves primarily as compensation for channel degradation due to the path loss and fading between the ground BS and the receiver UAV. Therefore, no additional diversity is achieved in this case.
    \item The relayed UAV utilizes an amplify-and-forward relaying protocol. Therefore, these results will serve as a lower bound when compared with the conventional decode-and-forward protocol.
\end{itemize}

\begin{figure}
\centering
    \begin{subfigure}[b]{0.45\textwidth}
        \centering
        \includegraphics[height=2.4in]{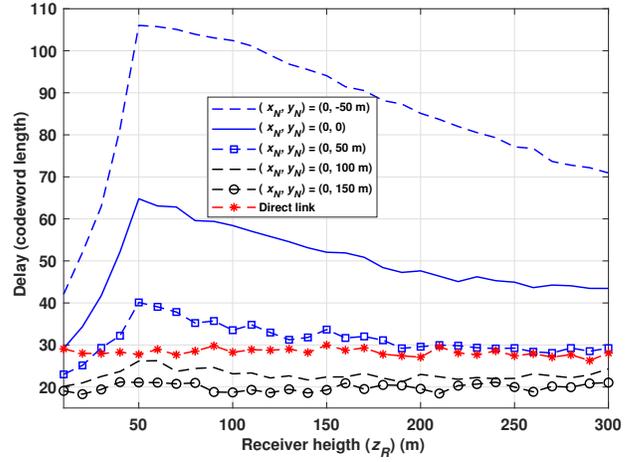}
        \caption{Received signal-to-noise power ratio $\lambda = 5$ dB.}
    \end{subfigure}
    
    \begin{subfigure}[b]{0.45\textwidth}
        \centering
        \includegraphics[height=2.4in]{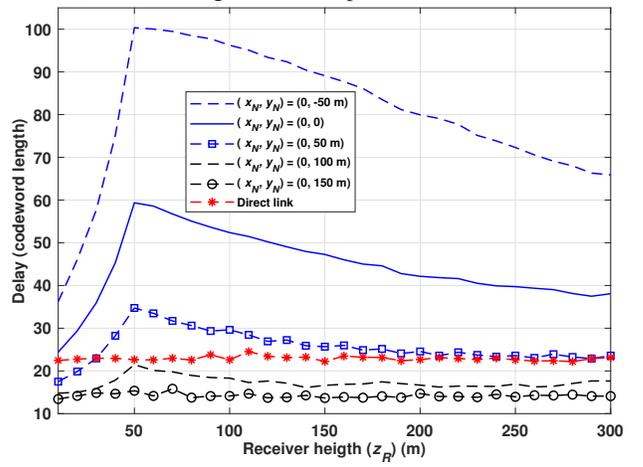}
        \caption{Received signal-to-noise power ratio $\lambda = 5$ dB.}
    \end{subfigure}
    
    \begin{subfigure}[b]{0.45\textwidth}
        \centering
        \includegraphics[height=2.4in]{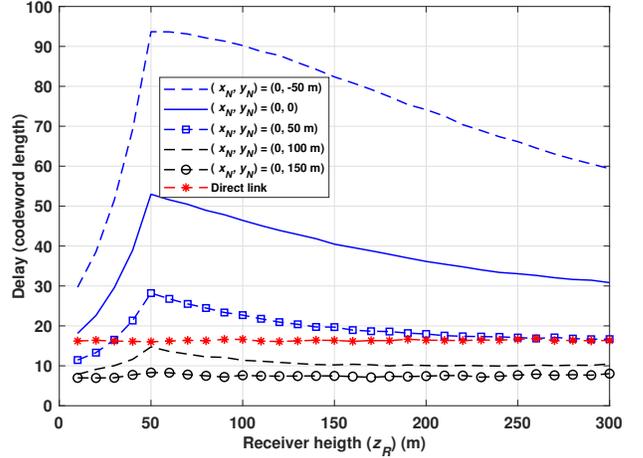}
        \caption{Received signal-to-noise power ratio $\lambda = 5$ dB.}
    \end{subfigure}
    \caption{Delay profile relative to the relay node location for different receiver heights (with noise amplification at the relay node).}
    \label{FIG-9}
\end{figure}

The results presented in Fig. \ref{FIG-9} are obtained with identical conditions as set in obtaining the results in Fig. \ref{FIG-8}, however, the relay node is assumed to have Gaussian noise at its input with noise amplification considered.

\begin{figure}
	\centering
	\includegraphics[width=0.47\textwidth]{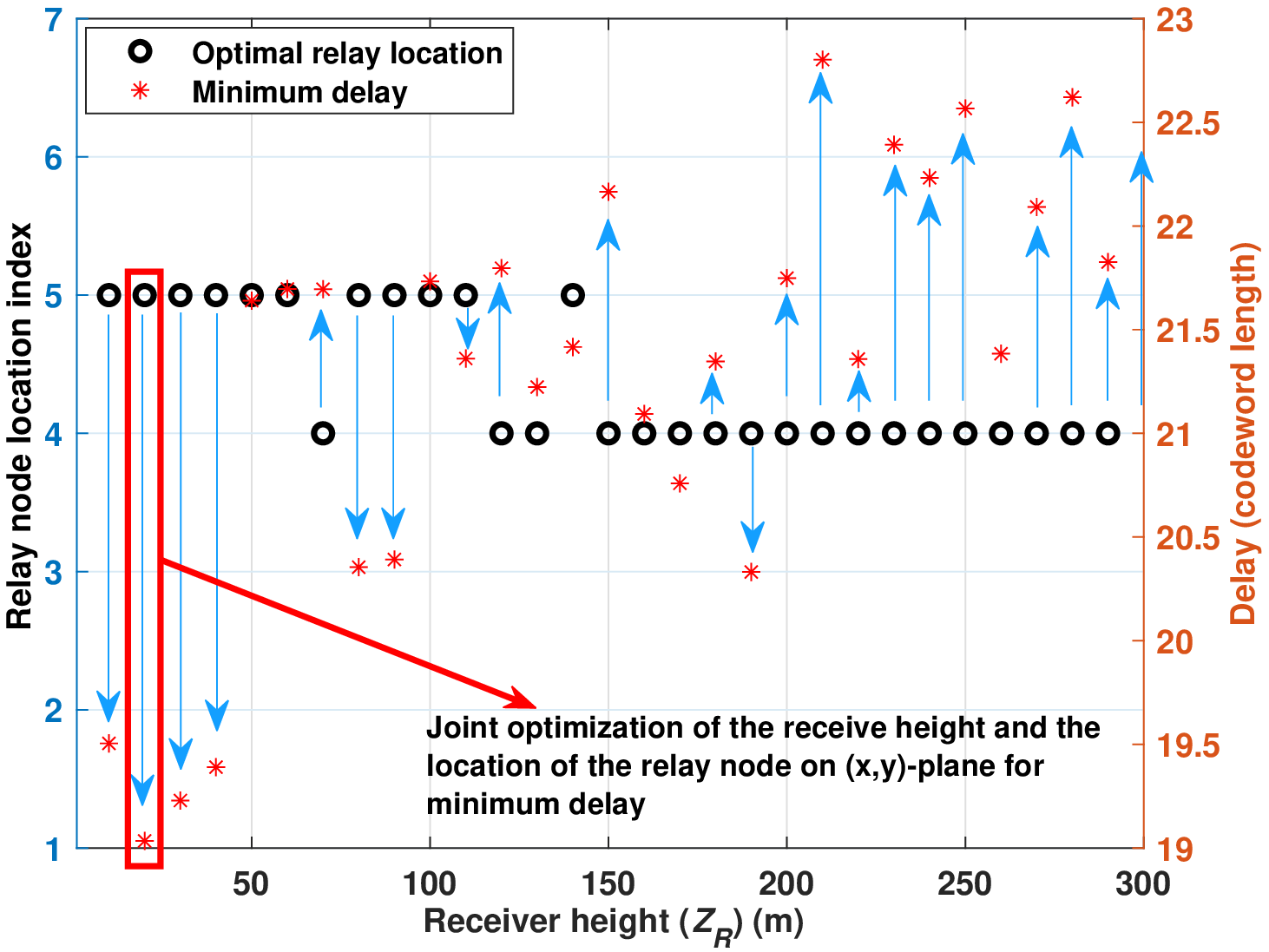}
	\caption{Optimal link configuration for minimum delay.}
	\label{FIG-10}
\end{figure}

\begin{figure}
	\centering
	\includegraphics[width=0.47\textwidth]{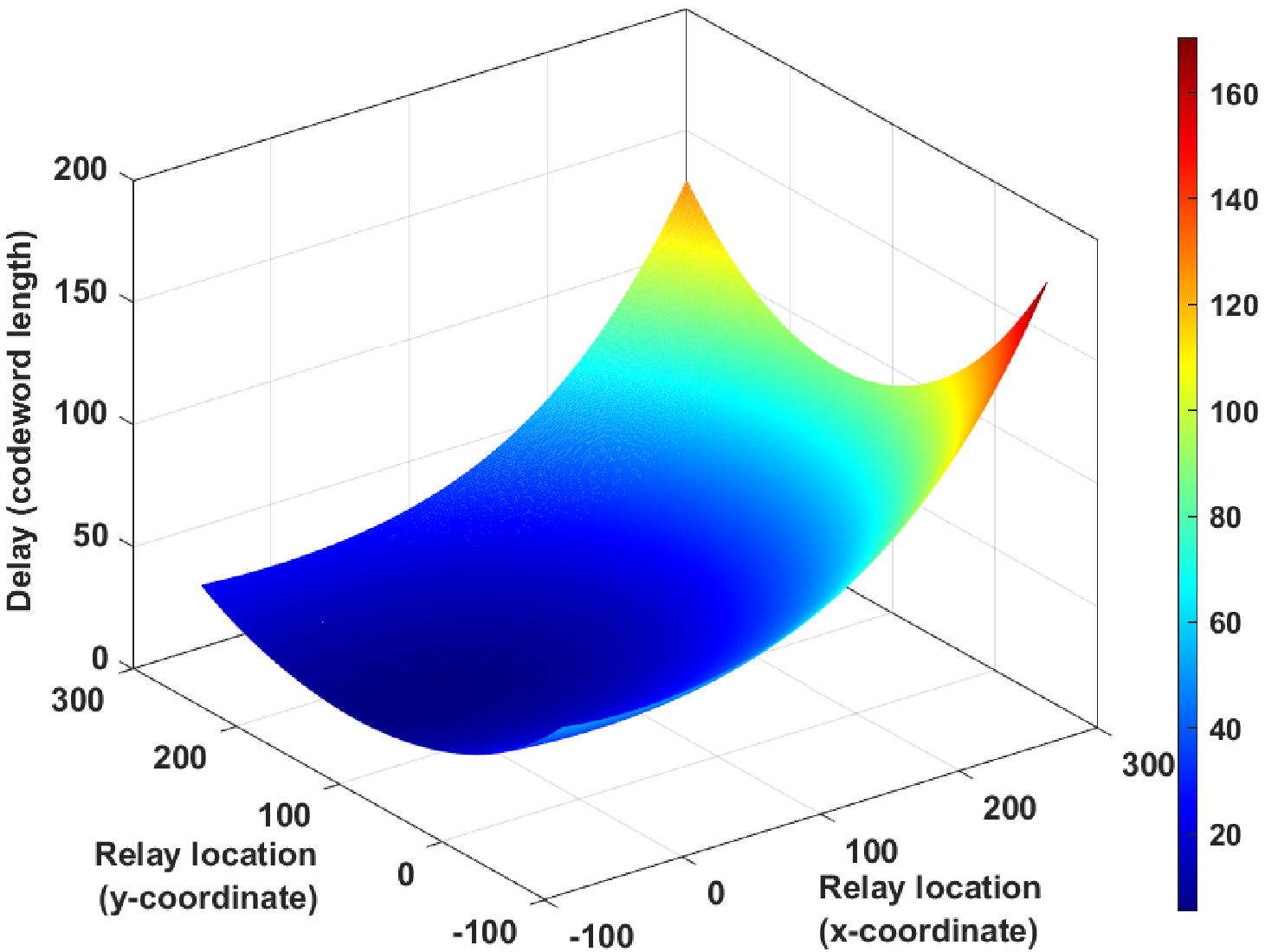}
	\caption{Delay profile relative to the relay node location.}
	\label{FIG-11}
\end{figure}

In Fig. \ref{FIG-10}, the joint optimization of the relay node and location and the receiver height is presented with the aim to minimize the delay. In obtaining the results in Fig. \ref{FIG-10}, the end-to-end received average signal-to-noise power ratio is set to $ - 2$ dB whereas the amplification gain of the relay node is set to $ -3$ dB. The relay node location index $\{1, 2, 3, 4, 5\}$ corresponds to the location coordinates as $(0, -50, 50)$, $(0, 00, 50)$, $(0, 50, 50)$, $(0, 100, 50)$, $(0, 150, 50)$, respectively. $x_R$ and $y_R$ are set to $0$ and $250$.

Figure \ref{FIG-11} shows the impact of the relay node location in minimizing the delay. Interestingly, it is to be noted that due to the hammock shape of the delay profile, as illustrated in Fig. \ref{FIG-11}, there is only one global minimum, and the optimal relay node location corresponding to this global minima can readily be obtained using convex optimization. Recall that in obtaining the results in Fig. \ref{FIG-11}, our dual-hop AF scheme does not exploit the existence of the direct link from the BS to the UAV for the coherent combining of signals at the receiver. Ignoring the direct link in this analysis provides an upper bound on the delay that can be achieved over an interference plus noise G2U channel.

\section{Conclusion}
In this work, we have presented an information-theoretic framework to study the delayed performance of the G2U integrated network. Utilizing Fano’s inequality, we have derived the tight upper bound for the channel capacity of the G2U network Based on the location information of the UAV for a point-to-point G2U, we have also obtained the tight lower bound on the transmit power requirement subject to the reliability constraint and the maximum delay threshold. Interestingly, it has been shown that, despite the simplicity of the point-to-point direct link, the latency of the AF relayed-based link can be lower, if the relay is properly placed. Our results show that increasing the transmit power may not always be an optimal solution for latency minimization problems, though it may significantly improve the outage performance. A possible extension of this work relates to investigating the ground-to-multi-relayed UAV networks, which can offer higher reliability at the cost of even more system complexity.

\appendix[Proof of (\ref{EQ-7})]\label{APPENDIX-1}
Utilizing (\ref{EQ-4}) and (\ref{EQ-5}) and applying the identities \cite[eq. 03.02.26.0006.01]{Wolfram} and \cite[eq. 01.03.26.0007.01]{Wolfram}, the average received SNR for the link $u \rightarrow v$ can be expressed as depicted in (\ref{EQ-A1}).

Applying the identity \cite[eq. 07.34.21.0009.01]{Wolfram}, and by utilizing the fact that the gamma function has simple poles at negative integers, it can be readily shown that the integral in the Term I in (\ref{EQ-A1}) is guaranteed to approach zero irrespective of the values of $\sigma_{h_{u,v}}^2$ and $\rho_{h_{u,v}}^2$. We like to point out that the gamma function is a meromorphic function that has simple poles at negative integers \cite{thukral2014factorials}. Finally, applying the identity \cite[eq. 07.34.21.0011.01]{Wolfram}, the integral depicted in Term II can be computed to a closed form as 
$2\left( { - \sigma _{h_{u,v} }^4 } \right)G_{3,4}^{2,1} \left[ {\frac{{2\rho _{h_{u,v} }^2 }}{{\sigma _{h_{u,v} }^2 }}\left| {\begin{array}{*{20}c}
   { - 2, - 1,\frac{1}{2}}  \\
   {0, - 2,0,\frac{1}{2}}  \\
\end{array}} \right.} \right]$. Substituting the equivalent closed-form expressions corresponding to Term I and Term II in (\ref{EQ-A1}), the expression for the average received SNR is obtained as shown in (\ref{EQ-7}).

\appendix[Proof of (\ref{EQ-15})]\label{APPENDIX-2}
Applying the chain rule and utilizing the fact that conditioning reduces entropy, $H\left( {Y_R\left| {s_I } \right.} \right)$ can readily be expressed as
\begin{equation}
    \begin{split}
 & H\left( {Y_R\left| {s_I } \right.} \right) \le \sum\limits_{j = 1}^{N_p/T_c} {\left\{ {\mathop {\max }\limits_{{\rm trace}\left[ {\mathbb{E}\left[ {s_{T,j} ,s_{T,j}^* } \right]} \right] \le T} \mathop \mathbb{E}\limits_{s_{I,j} } \left[ {\log \left( {\pi e} \right)^T } \right.} \right.}  \\ 
 & \hspace{0.4cm} \times \det \left[ {P_T d_{T,R}^{ - \alpha _{T,R} \left( {\theta _{T,R} } \right)} h_{T,R} \mathbb{E}\left[ {s_{T,j} ,s_{T,j}^* } \right]} \right. \\ 
 & \hspace{0.35cm} \left. {\left. {\left. { + \sum\limits_{i \in \Phi _I } {P_I^{\left( i \right)} d_{i,R}^{ - \alpha _{I,R}^{\left( i \right)} \left( {\theta _{I,R}^{\left( i \right)} } \right)} \left[ {s_{I,j}^{\left( i \right)} \left( {s_{I,j}^{\left( i \right)} } \right)^ *  } \right] + B_{TR}N_0 I} } \right]} \right]} \right\} \\
    \end{split}
    \label{EQ-B1}.
\end{equation}
where $I$ is the identity matrix. Subscript $j$ represents the block index. It follows that $N_p/T_c$ should be a positive integer. Following a similar approach as illustrated in \cite{zhang2015capacity}, with some simple mathematical manipulations, (\ref{EQ-B1}) can be written as illustrated in (\ref{EQ-B2}).

Utilizing the chain rule for entropy, $H\left( {Y\left| {S,s_I } \right.} \right)$ can readily be expressed as shown in (\ref{EQ-B3}). Applying the Markov chain, the expression in (\ref{EQ-B3}) is reduced to the expression as illustrated in (\ref{EQ-B4}).
\begin{figure*}[t]
    \begin{equation}
    \begin{split}
     &   H\left( {Y_R\left| {s_I } \right.} \right) \le N_p\log \left( {\pi e} \right) + \frac{{N_p\left( {T_c - 1} \right)}}{T_c}\log \left( {P_T d_{T,R}^{ - \alpha _{T,R} \left( {\theta _{T,R} } \right)}  + B_{TR}N_0 } \right) \\ 
 & + \sum\limits_{j = 1}^{N_p/T_c} {\mathop E\limits_{s_{I,j} } \left[ {\log \left( {P_T d_{T,R}^{ - \alpha _{T,R} \left( {\theta _{T,R} } \right)}  + \sum\limits_{i \in \Phi _I } {P_I^{\left( i \right)} d_{i,R}^{ - \alpha _{I,R}^{\left( i \right)} \left( {\theta _{I,R}^{\left( i \right)} } \right)} \left\| {s_{I,j}^{\left( i \right)} } \right\|^2  + B_{TR}N_0 } } \right)} \right]}  \\ 
    \end{split}
    \label{EQ-B2}.
     \end{equation}
     \begin{center}
    \line(1,0){380}
 \end{center}
\end{figure*}
    
\begin{figure*}[t]
    \begin{equation}
        \begin{split}
& H\left( {Y_R\left| {S,s_I } \right.} \right) = \sum\limits_{j = 1}^{N_p/T_c} {H\left( {y_j \left| {y_1 , \cdot  \cdot  \cdot ,y_{j - 1} ,S,s_I } \right.} \right)}  \\ 
&  \hspace{1.8cm}= \sum\limits_{j = 1}^{N_p/T_c} {H\left( {\left. {\sum\limits_{i \in \Phi _I } {\sqrt {P_I^{\left( i \right)} d_{I,R}^{ - \alpha _{I,R}^{\left( i \right)} \left( {\theta _{I,R}^{\left( i \right)} } \right)} } h_{I,j}^{\left( i \right)} s_{I,j}^{\left( i \right)} }  + w_R } \right|y_1 , \cdot  \cdot  \cdot ,y_{j - 1} ,S,s_T ,s_I } \right)}  \\
        \end{split}
        \label{EQ-B3}.
    \end{equation}
    \begin{center}
    \line(1,0){380}
 \end{center}
\end{figure*}

\begin{equation}
    \begin{split}
 &H\left( {Y\left| {S,s_I } \right.} \right) =  \\ 
 & \hspace{1cm}N_p\log \left( {\pi e} \right) + \frac{{N_p\left( {T_c - 1} \right)}}{T_c}\log \left( {B_{TR}N_0} \right) \\ 
  & \hspace{1cm} + \sum\limits_{j = 1}^{N_p/T_c} {\mathop \mathbb{E}\limits_{s_I ,j} \log \left( {\sum\limits_{i \in \Phi _I } {P_I^{\left( i \right)} d_{I,R}^{ - \alpha _{I,R}^{\left( i \right)} \left( {\theta _{I,R}^{\left( i \right)} } \right)}\left\| {s_{I,j}^{\left( i \right)} } \right\|} ^2  + w_R } \right)}  \\
    \end{split}
    \label{EQ-B4}.
\end{equation}
Substituting (\ref{EQ-B2}) and (\ref{EQ-B4}) into (\ref{EQ-10}), and applying simple mathematical manipulations, yields (\ref{EQ-15}).

\ifCLASSOPTIONcaptionsoff
  \newpage
\fi




\bibliographystyle{IEEEtran}


\begin{IEEEbiography}[{\includegraphics[width=1.0in,height=1.25in, clip,keepaspectratio]{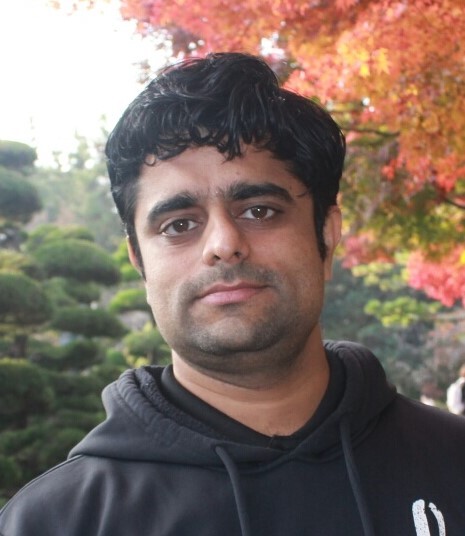}}]{Sudhanshu Arya} (Member, IEEE) is a researcher in the School of System and Enterprises at Stevens Institute of Technology, NJ, USA. He received his M.Tech. degree in communications and networks from the National Institute of Technology, Rourkela, India, in 2017, and the Ph.D. degree from Pukyong National University, Busan, South Korea, in 2022. He worked as a Research Associate with the Department of Artificial Intelligence Convergence, at Pukyong National University. His research interests include wireless communications and digital signal processing, with a focus on free-space optical communications, optical scattering communications, optical spectrum sensing, computational game theory, and artificial intelligence. He received the Best Paper Award in ICGHIT 2018 and the Early Career Researcher Award from the Pukyong National University in 2020.
\end{IEEEbiography}

\begin{IEEEbiography}[{\includegraphics[width=1in,height=1.25in, clip,keepaspectratio]{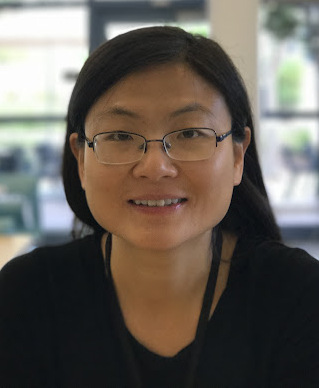}}]{Ying Wang} (Member, IEEE) received the B.E. degree in information engineering at Beijing University of Posts and Telecommunications, M.S. degree in electrical engineering from University of Cincinnati and the Ph.D. degree in electrical engineering from Virginia Polytechnic Institute and State University. She is an associate professor in the School of System and Enterprises at Stevens Institute of Technology. Her research areas include cybersecurity, wireless AI, edge computing, health informatics, and software engineering. 
\end{IEEEbiography}

\vfill

\end{document}